\documentclass[12pt]{article}
\usepackage{graphicx,color}
\usepackage{XoohmE}
\usepackage{booktabs}
\def\bo{{\raise.005ex\hbox{\large$\Box$}}}

 \def\kcr{{\hbox{\ro \char'170}}}                
 \def\ktl{{\hbox{\ro \char'170}}}        
 \def\ktr{{\hbox{\ro \char'170}}}        
 \def\kbl{{\hbox{\ro \char'170}}}        
 \def\kbr{{\hbox{\ro \char'170}}}        

 \parskip=\medskipamount                 
 \thicklines                         

 \def\bpl{\Big(}
 \def\bpr{\Big)}
 \def\e{\epsilon}
 \def\ve{\varepsilon}
 
 \def\der{\partial}
 \def\brr{\begin{equation}}
 \def\err{\end{equation}}
 \def\brr{\begin{eqnarray}}
 \def\err{\end{eqnarray}}
 \def\ba{\left(\begin{array}}
 \def\ea{\end{array}\right)}
 \def\lf{\left.\begin{array}{c}}
 \def\rf{\end{array}\right.}
 
 \def\dslash{\hbox{\ooalign{$\displaystyle\partial$\cr$/$}}}

 \newcommand{\dr}{\raise.3ex\hbox{$\stackrel{\leftarrow}{\partial }$}{}}
 \newcommand{\dl}{\raise.3ex\hbox{$\stackrel{\rightarrow}{\partial}$}{}}
 \newcommand{\topi}{\raise.3ex\hbox{$\stackrel{\pi}{\longrightarrow}$}{}}

 \def\qd{{\kern0.5pt
                   q \kern-5.05pt \raise5.8pt\hbox{$\textstyle.$}\kern 0.5pt}}

\def\4{\oplus}
\def\8{\otimes}

\definecolor{Hey}{rgb}{.9,.05,.4}
\definecolor{orange}{rgb}{1,.5,0}
\definecolor{plum}{rgb}{.4,0,.6}
\definecolor{R}{rgb}{1,0,0}
\definecolor{G}{rgb}{0,1,0}
\definecolor{B}{rgb}{0,0,1}
\font\cfnt=lcircle10 at 9pt
\def\lplus{\mathop{\kern2pt
            \raise1.275ex\hbox to0pt{\cfnt\char"07\hss}\kern-.6pt+}}
\def\YT#1#2{\vcenter{\hbox{\vbox{\baselineskip0pt\parskip=\medskipamount%
             \def\B{$\sqcap$\llap{$\sqcup$}\kern-1.9pt}
              \def\Bd{\hbox{\kern2.4pt\raise.4pt\hbox{$\cdot$}\kern-5.7pt\B\kern0pt}}
              \def\4{\raise.25pt\hbox to0pt{\hss\kern2pt--\hss}}
              \def\Z{\hfil\vskip-5.9pt}\lineskiplimit0pt\lineskip0pt%
               \setbox0=\hbox{#1}\hsize\wd0\parindent=0pt#2}\,}}}
\newdimen\parshift\parshift=\parindent
\catcode`@=11
 \long\def\@footnotetext#1{\insert\footins{\reset@font\footnotesize\interlinepenalty%
  \interfootnotelinepenalty\splittopskip\footnotesep\splitmaxdepth\dp\strutbox%
   \floatingpenalty\@MM\hsize\columnwidth\addtolength{\hsize}{-2\parindent}
    \@parboxrestore\protected@edef\@currentlabel{\csname p@footnote\endcsname\@thefnmark}
      \color@begingroup
       \@makefntext{\rule\z@\footnotesep\ignorespaces#1\@finalstrut\strutbox}
        \color@endgroup}}
 \long\def\@makefntext#1{\hglue\parshift
                         \vbox{\noindent\hb@xt@0em{\hss\@makefnmark\,}#1}}
\catcode`@=12
 \def\bpl{\Big(}
 \def\bpr{\Big)}
 \def\ba{\left(\begin{array}}
 \def\ea{\end{array}\right)}
 
 \def\der{\partial}
 \def\brr{\begin{eqnarray}}
 \def\err{\end{eqnarray}}

 \def\S{\mathbb S}

 \def\dslash{\hbox{\ooalign{$\displaystyle\partial$\cr$/$}}}
 \newcommand{\fr}[2]{{\textstyle\frac{#1}{#2}}}

 \HarvTitles 
 \seceq

\def\be{\begin{equation}}
\def\ee{\end{equation}}
\def\bea{\begin{eqnarray}}
\def\eea{\end{eqnarray}}

 \def\oldheadpic{                                
        \setlength{\unitlength}{.4mm}
        \thinlines
        \par
        \begin{picture}(349,16)
        \put(325,16){\line(1,0){4}}
        \put(330,16){\line(1,0){4}}
        \put(340,16){\line(1,0){4}}
        \put(335,0){\line(1,0){4}}
        \put(340,0){\line(1,0){4}}
        \put(345,0){\line(1,0){4}}
        \put(329,0){\line(0,1){16}}
        \put(330,0){\line(0,1){16}}
        \put(339,0){\line(0,1){16}}
        \put(340,0){\line(0,1){16}}
        \put(344,0){\line(0,1){16}}
        \put(345,0){\line(0,1){16}}
        \put(329,16){\oval(8,32)[bl]}
        \put(330,16){\oval(8,32)[br]}
        \put(339,0){\oval(8,32)[tl]}
        \put(345,0){\oval(8,32)[tr]}
        \end{picture}
        \par
        \thicklines
        \vskip.2in}
 \def\oldtitle#1#2#3#4{\oldheadpic\begin{center}\vglue.5in{\large\bf #1}\\[.6in]
        {#2}\\[.1in] {\it Department of Physics and Astronomy}\\
        {\it University of Maryland, College Park, MD 20742}\\[.6in]
        Physics Publication \#{#3}\\ {#4}\\[1.5in] {\bf ABSTRACT}\\[.1in]
        \end{center} \begin{quotation}}                 
 \def\oldTitle#1#2#3#4#5#6#7{\oldheadpic\begin{center} \vglue .4in
        {\large\bf #1}\\[.4in]
        {#2}\\[.1in] {\it Department of Physics and Astronomy}\\
        {\it University of Maryland, College Park, MD 20742}\\[.1in]
        {#3}\\[.1in] {\it {#4}}\\ {\it {#5}}\\[.4in]
        Physics Publication \#{#6}\\ {#7}\\[.5in] {\bf ABSTRACT}\\[.1in]
        \end{center} \begin{quotation}}                 
 \font\rOpe=cmsy10                        
 \def\ktl{{\hbox{\rOpe\char'170}}}        
 \def\kbl{{\hbox{\rOpe\char'170}}}        
 \def\kcr{{\reflectbox{\rOpe\char'170}}}        
 \def\ktr{{\reflectbox{\rOpe\char'170}}}        
 \def\kbr{{\reflectbox{\rOpe\char'170}}}        
 \def\Border{\vbox{\hsize0pt
        \setlength{\unitlength}{1mm}
        \newcount\xco
        \newcount\yco
        \xco=-21
        \yco=12
        \begin{picture}(0,0)(6.5,-5)
        \put(\xco,\yco){$\ktl$}
        \advance\yco by-1
        {\loop
        \put(\xco,\yco){$\kcr$}
        \advance\yco by-2
        \ifnum\yco>-240
        \repeat
        \put(\xco,\yco){$\kbl$}}
        \xco=170
        \yco=12
        \put(\xco,\yco){$\ktr$}
        \advance\yco by-1
        {\loop
        \put(\xco,\yco){$\kcr$}
        \advance\yco by-2
        \ifnum\yco>-240
        \repeat
        \put(\xco,\yco){$\kbr$}}
        \put(-19.5,13){\scalebox{.598}{- SUNY College at Oneonta
            Physics Department|University of Maryland Center for
            String and Particle  Theory \&\ Physics Department|%
            Delaware State University DAMTP -}}
        \put(-19.5,-241.5){\scalebox{.728}{University of Washington
            Mathematics Department|Pepperdine University Natural
            Sciences Division|Bard College Mathematics
            Department}}
        \end{picture}
        \par\vskip-8mm}}
\definecolor{UMred}{rgb}{.9,.05,.2}
 \def\UMbanner{\vbox{\hsize0pt
        \setlength{\unitlength}{.4mm}
        \thicklines
        \begin{picture}(0,0)(7,-10)
        \put(165,16){\line(1,0){4}}
        \put(170,16){\line(1,0){4}}
        \put(180,16){\line(1,0){4}}
        \put(175,0){\line(1,0){4}}
        \put(180,0){\line(1,0){4}}
        \put(185,0){\line(1,0){4}}
        \put(169,0){\line(0,1){16}}
        \put(170,0){\line(0,1){16}}
        \put(179,0){\line(0,1){16}}
        \put(180,0){\line(0,1){16}}
        \put(184,0){\line(0,1){16}}
        \put(185,0){\line(0,1){16}}
        \put(169,16){\oval(8,32)[bl]}
        \put(170,16){\oval(8,32)[br]}
        \put(179,0){\oval(8,32)[tl]}
        \put(185,0){\oval(8,32)[tr]}
        \end{picture}
        \par\vskip-6.5mm
        \thicklines}}

 \def\qd{{\kern0.5pt
                   q \kern-5.05pt \raise5.8pt\hbox{$\textstyle.$}\kern 0.5pt}}

\begin{document}
\thispagestyle{empty} \vbox{\Border\UMbanner} \noindent{October
2007}\hfill {UMDEPP 07-011, SUNY/O-666}

 \vspace{.4in}

 \setlength{\oddsidemargin}{0.3in}
 \setlength{\evensidemargin}{-0.3in}
 \begin{center}
 \vglue .05in {\Large\bf On the Matter of ${\cal N} =2$ Matter}
 \\[.25in]

  { C.F.\,Doran$^a$, M.G.\,Faux$^b$, S.J.\,Gates, Jr.$^c$,
     T.\,H\"{u}bsch$^d$, K.M.\,Iga$^e$ and G.D.\,Landweber$^f$}\\[3mm]
{\small\it
  $^a$Department of Mathematics,
      University of Washington, Seattle, WA 98105\\[-1mm]
  {\tt  doran@math.washington.edu}
  \\
  $^b$Department of Physics,
      State University of New York, Oneonta, NY 13820\\[-1mm]
  {\tt  fauxmg@oneonta.edu}
  \\
  $^c$Department of Physics,
      University of Maryland, College Park, MD 20472\\[-1mm]
  {\tt  gatess@wam.umd.edu}
  \\
  $^d$Department of Physics and Astronomy,
      Howard University, Washington, DC 20059\\
      DAMTP, Delaware State University, Dover, DE 19901\\[-1mm]
  {\tt  thubsch@desu.edu}
  \\
  $^e$Natural Science Division,
      Pepperdine University, Malibu, CA 90263\\[-1mm]
  {\tt  Kevin.Iga@pepperdine.edu}
  \\
 $^f$Department of Mathematics, Bard College, Annandale-on-Hudson, NY
 12504-5000\\[-1mm]
  {\tt gregland@bard.edu}
 }\\[6mm]

 {\bf ABSTRACT}\\[.01in]
 \end{center}
 \begin{quotation}
 {We introduce a variety of four-dimensional ${\cal N} =2$ matter multiplets which
 have not previously appeared explicitly in the literature.  Using these, we develop
 a class of supersymmetric actions supplying a context for a systematic
 exploration of ${\cal N} =2$ matter theories, some of which include
 Hypermultiplet sectors in novel ways.  We construct an ${\cal N} =2$ supersymmetric
 field theory in which the propagating fields are realized off-shell exclusively as
 Lorentz scalars and Weyl spinors and which involves a sector with
 precisely the $R$-charge assignments characteristic of
 Hypermultiplets.}

 ${~~~}$ \newline PACS: 04.65.+e

\end{quotation}\clearpage\setcounter{page}{1}

  There are various ways to realize four-dimensional ${\cal N} =2$
 Hypermultiplets \cite{Fayet} off-shell. The known examples
 have a nontrivial off-shell central charge \cite{Sohnius}, or have infinite
 independent degrees of freedom \cite{Harmonic}, or involve constrained
 vector fields \cite{Tensors}.  It is a relevant and interesting open question whether
 a Hypermultiplet with no off-shell central
 charge, finite off-shell degrees of freedom, and no
 constrained vector fields, exists.  In this letter we introduce
 several ${\cal N} =2$ matter multiplets, and a variety of corresponding supersymmetric
 actions,
 which facilitate a systematic exploration of this prospect.
 By way of partial solution, we construct an ${\cal N} =2$ supersymmetric
 field theory in which the propagating fields are realized off-shell exclusively as
 Lorentz scalars and Weyl spinors and which involves a sector with
 precisely the $R$-charge assignments characteristic of
 Hypermultiplets.

 In our discussion, a central role is played by constrained ${\cal N} =2$ superfields which transform as
 symmetric rank-$p$ tensor products of fundamental $SU(2)_R$ representations. The cases $p=2, 3, 4$
 and $5$ describe a Triplet, a Quadruplet, a Quintet, and a Sextet,
 respectively, and we refer to the corresponding multiplets using
 these names.
 The well-known ${\cal N} =2$ Tensor multiplet corresponds to a constrained Triplet
 superfield,
 and includes as off-shell components a dimension-one
 boson triplet, a dimension-three-half Weyl spinor doublet, and
 five dimension-two bosons transforming as a complex Lorentz scalar and a
 divergence-free Lorentz vector, the latter of which describes the
 Hodge dual of the three-form field strength associated with a
 two-form gauge potential.

 In an early attempt to realize a Hypermultiplet off-shell as
 a ``relaxed" version of a Tensor multiplet, the authors of
 \cite{HST} utilized the highest component of the Quintet
 superfield---a dimension-three real scalar boson---to supply the divergence of the erstwhile
 constrained vector component within the
 Tensor multiplet.  A Dual Quintet superfield, which is a
 constrained scalar superfield admitting a supersymmetric
 coupling ({\it i.e.} an inner product) with the Quintet superfield, provides
 Lagrange multiplier terms which render the five dimension-one bosons within
 the Quintet multiplet non-propagating.  What results is a supersymmetric field theory
 reproducing the on-shell state counting characteristic of
 Hypermultiplets, but with $R$-charge assignments which differ.

 The four propagating bosons in a Hypermultiplet transform as a doublet
 under $SU(2)_R$ and also as a doublet under an additional
 symmetry $SU(2)_H$.  The four propagating fermions assemble as a
 pair of Weyl spinors which transform as $SU(2)_R$ singlets and $SU(2)_H$
 doublets. This structure admits a well-known generalization whereby
 $SU(2)_H$ is replaced with $Sp(2\,n)$ for $n\in {\mathbb N}$.  A natural way
 to realize component fields with these representation assignments is obtained by
 starting with an ${\cal N} =2$ Tensor multiplet and augmenting this by
 ``tensoring" two additional indices, one corresponding to the
 fundamental $SU(2)_R$ representation and another to
 the fundamental $SU(2)_H$ representation.  When the
 component fields are expressed in terms of irreducible
 representations of $SU(2)_R$, one obtains a standard presentment of
 what we call the Extended Tensor Multiplet.

 To be more explicit, the Tensor
 multiplet is specified by the superfield ${\mathbb L}_{ij}={\mathbb L}_{(ij)}$,
 subject to the constraint $D_{\alpha\,(i}{\mathbb
 L}_{jk)}=0$, where $D_{\alpha\,i}$ is a superspace derivative, and
 to the reality constraint ${\mathbb
 L}_{ij}=\ve_{im}\,\ve_{jn}\,{\mathbb L}^{mn}$, where complex
 conjugation is indicated by the raising or lowering of $SU(2)$
 indices.  The lowest components of the Tensor multiplet comprise a boson
 triplet $L_{ij}$.  The lowest components of the Extended Tensor multiplet comprise the ``tensored" analog
 $L_{ij}\,^{k\,\hat{\imath}}$, where the hatted index corresponds to
 the fundamental representation of $SU(2)_H$, subject to
 the reality constraint
 $L_{ij}\,^{p\,\hat{\imath}}=\ve_{im}\,\ve_{jn}\,\ve^{pq}\,\ve^{\hat{\imath}\hat{\jmath}}\,L^{mn}\,_{q\hat{\jmath}}$.
 We decompose $L_{ij}\,^{p\hat{\imath}}$ into
 $SU(2)_R$ irreps as
 \brr L_{ij}\,^{p\hat{\imath}} &=&
      \fr23\,\delta_{(i}\,^p\,\phi_{j)}\,^{\hat{\imath}}
      +\ve^{pk}\,u_{ijk}\,^{\hat{\imath}} \,,
 \err
 where $\phi_i\,^{\hat{\imath}}$ is a doublet-doublet, and
 $u_{ijk}\,^{\hat{\imath}}$ is a quadruplet-doublet under
 $SU(2)_R\times SU(2)_H$.  These fields are subject to the
 constraints
 $\phi_i\,^{\hat{\imath}}=
 \ve_{ij}\,\ve^{\hat{\imath}\hat{\jmath}}\,\phi^j\,_{\hat{\jmath}}$
 and
$u_{ijk}\,^{\hat{\imath}}=
\ve_{il}\,\ve_{jm}\,\ve_{kn}\,\ve^{\hat{\imath}\hat{\jmath}}\,u^{lmn}\,_{\hat{\jmath}}$.
 It is straightforward to determine the supersymmetry transformation
 rules for a complete supermultiplet which utilizes
 $\phi_i\,^{\hat{\imath}}$ and $u_{ijk}\,^{\hat{\imath}}$ as its
 lowest components.  These are given by
  \brr \delta_Q\,\phi_i\,^{\hat{\imath}} &=&
      i\,\bar{\e}_i\,\psi^{\hat{\imath}}
      +i\,\ve^{mn}\,\bar{\e}_m\,
      \lambda_{ni}\,^{\hat{\imath}}
      +i\,\ve_{ij}\,\ve^{\hat{\imath}\hat{\jmath}}\,\bar{\e}^j\,\psi_{\hat{\jmath}}
      +i\,\ve_{ij}\,\ve_{mn}\,\ve^{\hat{\imath}\hat{\jmath}}\,\bar{\e}^m\,\lambda^{nj}\,_{\hat{\jmath}}
      \nonumber\\[.1in]
      \delta_Q\,u_{ijk}\,^{\hat{\imath}} &=&
      i\,\bar{\e}_{(i}\,\lambda_{jk)}\,^{\hat{\imath}}
      +i\,\ve_{im}\,\ve_{jn}\,\ve_{kp}\,\ve^{\hat{\imath}\hat{\jmath}}\,\bar{\e}^{(m}\,\lambda^{np)}\,_{\hat{\jmath}}
      \nonumber\\[.1in]
      \delta_Q\,\psi^{\hat{\imath}} &=&
      \fr32\,\dslash\,\phi_i\,^{\hat{\imath}}\,\e^i
      +A_{a\,i}\,^{\hat{\imath}}\,\gamma^a\,\e^i
      +\ve^{mn}\,N_m\,^{\hat{\imath}}\,\e_n
      \nonumber\\[.1in]
      \delta_Q\,\lambda_{ij}\,^{\hat{\imath}} &=&
      \fr13\,\ve_{k(i}\,\dslash\,\phi_{j)}\,^{\hat{\imath}}\,\e^k
      +2\,\dslash\,u_{ijk}\,^{\hat{\imath}}\,\e^k
      -\fr23\,\ve_{k(i}\,A_{a\,j)}\,^{\hat{\imath}}\,\gamma^a\,\e^k
      -\fr23\,N_{(i}\,^{\hat{\imath}}\,\e_{j)}
      \nonumber\\[.1in]
      \delta_Q\,N_i\,^{\hat{\imath}} &=&
      i\,\ve_{ij}\,\bar{\e}^j\,\dslash\,\psi^{\hat{\imath}}
      -3\,i\,\bar{\e}^j\,\dslash\,\lambda_{ij}\,^{\hat{\imath}}
      \nonumber\\[.1in]
      \delta_Q\,A_{a\,i}\,^{\hat{\imath}} &=&
      \fr12\,i\,\bar{\e}_i\,\gamma_{ab}\,\der^b\psi^{\hat{\imath}}
      -\fr32\,i\,\ve^{mn}\,\bar{\e}_m\,\gamma_{ab}\,\der^b\lambda_{ni}\,^{\hat{\imath}}
      \nonumber\\[.1in]
      & &
      +\fr12\,i\,\ve_{ij}\,\ve^{\hat{\imath}\hat{\jmath}}\,\bar{\e}^j\,\gamma_{ab}\,\der^b\psi_{\hat{\jmath}}
      -\fr32\,i\,\ve_{ij}\,\ve_{mn}\,\ve^{\hat{\imath}\hat{\jmath}}\,\bar{\e}^m\,\gamma_{ab}\,\der^b
      \lambda^{nj}\,_{\hat{\jmath}} \,,
 \label{etet}\err
 where $\psi^{\hat{\imath}}$ is an $SU(2)_H$ doublet of
 Weyl spinors, $\lambda_{ij}\,^{\hat{\imath}}=\lambda_{(ij)}\,^{\hat{\imath}}$ is an assembly of six
 Weyl spinors transforming as a triplet-doublet under $SU(2)_R\times
 SU(2)_H$, $N_i\,^{\hat{\imath}}$ describes four complex bosons
 transforming as a doublet-doublet, and $A_{a\,i}\,^{\hat{\imath}}$
 is a Lorentz vector transforming as a
 doublet-doublet, subject to the reality constraint
 $A_{a\,i}\,^{\hat{\imath}}=\ve_{ij}\,\ve^{\hat{\imath}\hat{\jmath}}\,A_a{}^j\,_{\hat{\jmath}}$,
 and also constrained to be divergence-free,
 $\der^aA_{a\,i}\,^{\hat{\imath}}=0$.
 The Extended Tensor transformation rules (\ref{etet}) describe 32+32 off-shell degrees of freedom,
 and represent the ${\cal N}=2$ susy algebra
 with no central charge.

 An action functional invariant under (\ref{etet}) is given by
 \brr S_{\rm ET} &=& \int d^4x\,\bpl
      -\fr34\,\phi_i\,^{\hat{\imath}}\,\Box\,\phi^i\,_{\hat{\imath}}
      -\fr92\,u_{ijk}\,^{\hat{\imath}}\,\Box\,u^{ijk}\,_{\hat{\imath}}
      +N_i\,^{\hat{\imath}}\,N^i\,_{\hat{\imath}}
      -A_{a\,i}\,^{\hat{\imath}}\,A^{a\,i}\,_{\hat{\imath}}
      \nonumber\\[.1in]
      & &
      \hspace{.7in}
      -i\,\bar{\psi}^{\hat{\imath}}\dslash \psi_{\hat{\imath}}
      -\fr92\,i\,\bar{\lambda}_{ij}\,^{\hat{\imath}}\,\dslash\lambda^{ij}\,_{\hat{\imath}}
      \,\bpr \,.
 \label{extenac}\err
 This model describes 16+16 propagating degrees of freedom,
 including the ``Hypermultiplet" fields $\phi_i\,^{\hat{\imath}}$
 and $\psi^{\hat{\imath}}$, but also includes extra matter fields
 $u_{ijk}\,^{\hat{\imath}}$, $A_{a\,i}\,^{\hat{\imath}}$, and
 $\lambda_{ij}\,^{\hat{\imath}}$.
 In order to realize an off-shell Hypermultiplet
 conforming
 to our specified criteria, we require a mechanism to ``relax" the
 divergence-free constraint imposed on $A_{a\,i}\,^{\hat{\imath}}$ and also to
 render this field auxiliary.  Thus, we seek another supermultiplet which
 can properly supply the divergence
 $\der^aA_{a\,i}\,^{\hat{\imath}}$.

 A suitable multiplet for relaxing the Extended Tensor multiplet should have as its highest components
 four dimension-two Lorentz scalar bosons
 transforming under $SU(2)_R\times SU(2)_H$ as a doublet-doublet and subject to a reality constraint
 $X_i\,^{\hat{\imath}}=\ve_{ij}\,\ve^{\hat{\imath}\hat{\jmath}}\,X^j\,_{\hat{\jmath}}$.   This requirement is fulfilled
 by the Sextet multiplet, which describes 64+64 off-shell degrees of
 freedom, and
 has the following component transformation rules,
    \brr \delta_Q\,Z_{ijklm}\,^{\hat{\imath}} &=&
      i\,\ve_{n(i}\,\bar{\e}^n\,\Sigma_{jklm)}\,^{\hat{\imath}}
      +i\,\ve_{ip}\,\ve_{jq}\,\ve_{kr}\,\ve_{ls}\,\ve_{mt}\,\ve^{\hat{\imath}\hat{\jmath}}\,\ve^{n(p}\,\bar{\e}_n\,
      \Sigma^{qrst)}\,_{\hat{\jmath}}
      \nonumber\\[.1in]
      \delta_Q\,\Sigma_{ijkl}\,^{\hat{\imath}} &=&
      -2\,\ve^{mn}\,\dslash\,Z_{ijklm}\,^{\hat{\imath}}\,\e_n
      +K_{a\,(ijk}\,^{\hat{\imath}}\,\gamma^a\,\e_{l)}
      +\ve_{m(i}\,P_{jkl)}\,^{\hat{\imath}}\,\e^m
      \nonumber\\[.1in]
      \delta_Q\,P_{ijk}\,^{\hat{\imath}} &=&
      2\,i\,\ve^{mn}\,\bar{\e}_m\,\dslash\,\Sigma_{nijk}\,^{\hat{\imath}}
      +i\,\bar{\e}_{(i}\,\xi_{jk)}\,^{\hat{\imath}}
      \nonumber\\[.1in]
      \delta_Q\,K_{a\,ijk}\,^{\hat{\imath}} &=&
      -\fr15\,i\,\bar{\e}^m\,(\,5\,\dslash\gamma_a+2\,\der_a\,)\,\Sigma_{mijk}\,^{\hat{\imath}}
      -\fr12\,i\,\ve_{m(i}\,\bar{\e}^m\,\gamma_a\,\xi_{jk)}\,^{\hat{\imath}}
      \nonumber\\[.1in]
      & &
      -\fr15\,i\,\ve_{ip}\,\ve_{jq}\,\ve_{kr}\,\ve^{\hat{\imath}\hat{\jmath}}\,\bar{\e}_m\,
      (\,5\,\dslash\gamma_a+2\,\der_a\,)\,\Sigma^{mpqr}\,_{\hat{\jmath}}
       \nonumber\\[.1in]
      & &
      -\fr12\,i\,\ve_{ip}\,\ve_{jq}\,\ve_{kr}\,\ve^{\hat{\imath}\hat{\jmath}}\,\ve^{m(p}\,
      \bar{\e}_m\,\gamma_a\,\xi^{qr)}\,_{\hat{\jmath}}
      \nonumber\\[.1in]
      \delta_Q\,\xi_{ij}\,^{\hat{\imath}} &=&
      \fr32\,\dslash P_{ijm}\,^{\hat{\imath}}\,\e^m
      +\fr12\,\ve^{mn}\,(\,3\,\dslash\gamma^a+8\,\der^a\,)\,K_{a\,ijm}\,^{\hat{\imath}}\,\e_n
      +X_{(i}\,^{\hat{\imath}}\,\e_{j)}
      \nonumber\\[.1in]
      \delta_Q\,X_i\,^{\hat{\imath}} &=&
      \fr43\,i\,\bar{\e}^m\,\dslash\,\xi_{mi}\,^{\hat{\imath}}
      +\fr43\,i\,\ve_{ij}\,\ve^{\hat{\imath}\hat{\jmath}}\,\bar{\e}_m\,\dslash\,\xi^{mj}\,_{\hat{\jmath}}
      \,.
 \label{sextet1}\err
 As a rule, adjacent $SU(2)_R$ indices are always symmetrized.
 For example, the lowest component $Z_{ijklm}\,^{\hat{\imath}}=Z_{(ijklm)}\,^{\hat{\imath}}$
 transforms as a sextet-doublet under $SU(2)_R\times SU(2)_H$.
 All bosons except $P_{ijk}\,^{\hat{\imath}}$ satisfy reality
 constraints, namely
 $X_i\,^{\hat{\imath}}=\ve_{ij}\,\ve^{\hat{\imath}\hat{\jmath}}\,X^j\,_{\hat{\jmath}}$,
  $K_{a\,ijk}\,^{\hat{\imath}}=\ve_{il}\,\ve_{jm}\,\ve_{kn}\,^{\hat{\imath}\hat{\jmath}}\,K_a^{lmn}\,_{\hat{\jmath}}$,
 and $Z_{ijklm}\,^{\hat{\imath}}=\ve_{in}\,\ve_{jp}\,\ve_{kq}\,\ve_{lr}\,\ve_{ms}\,^{\hat{\imath}\hat{\jmath}}\,
 Z^{npqrs}\,_{\hat{\jmath}}$.

  We can ``relax" the Extended Tensor multiplet using the Sextet
 multiplet, by identifying the component $X_i\,^{\hat{\imath}}$
 with $\der^aA_{a\,i}\,^{\hat{\imath}}$.   Under this circumstance the supersymmetry algebra
 closes on the component fields only if the rule $\delta_Q\,A_{a\,i}\,^{\hat{\imath}}$ is
 augmented by $\xi_{ij}\,^{\hat{\imath}}$ dependent terms.
 These modifications are easy to determine from the transformation rule
 $\delta_Q\,\xi_{ij}\,^{\hat{\imath}}$.  Furthermore, the
 commutator $\delta^2_Q\,N_i\,^{\hat{\imath}}$ is proportional
 to $\der^aA_{a\,i}\,^{\hat{\imath}}$.  Since this is no longer zero,
 this indicates the need for a
 $\xi_{ij}\,^{\hat{\imath}}$ dependent term to be added to
 $\delta_Q\,N_i\,^{\hat{\imath}}$.  This process continues to other components,
 and can be resolved completely.
 The fully amended transformation rules for the Relaxed
 Extended Tensor multiplet, which describes 96+96 component degrees of freedom, are
  \brr \delta_Q\,\phi_i\,^{\hat{\imath}} &=&
      i\,\bar{\e}_i\,\psi^{\hat{\imath}}
      +i\,\ve^{mn}\,\bar{\e}_m\,
      \lambda_{ni}\,^{\hat{\imath}}
      +i\,\ve_{ij}\,\ve^{\hat{\imath}\hat{\jmath}}\,\bar{\e}^j\,\psi_{\hat{\jmath}}
      +i\,\ve_{ij}\,\ve_{mn}\,\ve^{\hat{\imath}\hat{\jmath}}\,\bar{\e}^m\,\lambda^{nj}\,_{\hat{\jmath}}
      \nonumber\\[.1in]
      \delta_Q\,u_{ijk}\,^{\hat{\imath}} &=&
      i\,\bar{\e}_{(i}\,\lambda_{jk)}\,^{\hat{\imath}}
      +i\,\ve_{im}\,\ve_{jn}\,\ve_{kp}\,\ve^{\hat{\imath}\hat{\jmath}}\,\bar{\e}^{(m}\,\lambda^{np)}\,_{\hat{\jmath}}
       \nonumber\\[.1in]
      & &
      -\fr{16}{15}\,i\,\bar{\e}^l\,\Sigma_{ijkl}\,^{\hat{\imath}}
      -\fr{16}{15}\,i\,\ve_{im}\,\ve_{jn}\,\ve_{kl}\,\ve^{\hat{\imath}\hat{\jmath}}\,\bar{\e}_r\,
      \Sigma^{rmnl}\,_{\hat{\jmath}}
      \nonumber\\[.1in]
      \delta_Q\,Z_{ijklm}\,^{\hat{\imath}} &=&
      i\,\ve_{n(i}\,\bar{\e}^n\,\Sigma_{jklm)}\,^{\hat{\imath}}
      +i\,\ve_{ip}\,\ve_{jq}\,\ve_{kr}\,\ve_{ls}\,\ve_{mt}\,\ve^{\hat{\imath}\hat{\jmath}}\,\ve^{n(p}\,\bar{\e}_n\,
      \Sigma^{qrst)}\,_{\hat{\jmath}}
      \nonumber\\[.1in]
      \delta_Q\,\psi^{\hat{\imath}} &=&
      \fr32\,\dslash\,\phi_i\,^{\hat{\imath}}\,\e^i
      +A_{a\,i}\,^{\hat{\imath}}\,\gamma^a\,\e^i
      +\ve^{mn}\,N_m\,^{\hat{\imath}}\,\e_n
      \nonumber\\[.1in]
      \delta_Q\,\lambda_{ij}\,^{\hat{\imath}} &=&
      \fr13\,\ve_{k(i}\,\dslash\,\phi_{j)}\,^{\hat{\imath}}\,\e^k
      +2\,\dslash\,u_{ijk}\,^{\hat{\imath}}\,\e^k
      -\fr23\,\ve_{k(i}\,A_{a\,j)}\,^{\hat{\imath}}\,\gamma^a\,\e^k
      -\fr23\,N_{(i}\,^{\hat{\imath}}\,\e_{j)}
      \nonumber\\[.1in]
      & &
      -\fr43\,\ve_{im}\,\ve_{jn}\,\ve^{\hat{\imath}\hat{\jmath}}\,P^{mnl}\,_{\hat{\jmath}}\,\e_l
      +\fr43\,K_{a\,ijm}\,^{\hat{\imath}}\,\gamma^a\,\e^m
      \nonumber\\[.1in]
      \delta_Q\,\Sigma_{ijkl}\,^{\hat{\imath}} &=&
      -2\,\ve^{mn}\,\dslash\,Z_{ijklm}\,^{\hat{\imath}}\,\e_n
      +K_{a\,(ijk}\,^{\hat{\imath}}\,\gamma^a\,\e_{l)}
      +\ve_{m(i}\,P_{jkl)}\,^{\hat{\imath}}\,\e^m
      \nonumber\\[.1in]
      \delta_Q\,N_i\,^{\hat{\imath}} &=&
      i\,\ve_{ij}\,\bar{\e}^j\,\dslash\,\psi^{\hat{\imath}}
      -3\,i\,\bar{\e}^j\,\dslash\,\lambda_{ij}\,^{\hat{\imath}}
      -\fr83\,i\,\ve_{ij}\,\ve_{mn}\,\ve^{\hat{\imath}\hat{\jmath}}\,\bar{\e}^m\,\xi^{nj}\,_{\hat{\jmath}}
      \nonumber\\[.1in]
      \delta_Q\,A_{a\,i}\,^{\hat{\imath}} &=&
      \fr12\,i\,\bar{\e}_i\,\gamma_{ab}\,\der^b\psi^{\hat{\imath}}
      -\fr32\,i\,\ve^{mn}\,\bar{\e}_m\,\gamma_{ab}\,\der^b\lambda_{ni}\,^{\hat{\imath}}
       \nonumber\\[.1in]
      & &
      +\fr43\,i\,\bar{\e}^m\,\gamma_a\,\xi_{mi}\,^{\hat{\imath}}
      +\fr12\,i\,\ve_{ij}\,\ve^{\hat{\imath}\hat{\jmath}}\,\bar{\e}^j\,\gamma_{ab}\,\der^b\psi_{\hat{\jmath}}
      -\fr32\,i\,\ve_{ij}\,\ve_{mn}\,\ve^{\hat{\imath}\hat{\jmath}}\,\bar{\e}^m\,\gamma_{ab}\,\der^b
      \lambda^{nj}\,_{\hat{\jmath}}
      \nonumber\\[.1in]
      & &
      +\fr43\,i\,\ve_{ij}\,\ve^{\hat{\imath}\hat{\jmath}}\,\bar{\e}_m\,\gamma_a\,\xi^{mj}\,_{\hat{\jmath}}
      \nonumber\\[.1in]
      \delta_Q\,P_{ijk}\,^{\hat{\imath}} &=&
      2\,i\,\ve^{mn}\,\bar{\e}_m\,\dslash\,\Sigma_{nijk}\,^{\hat{\imath}}
      +i\,\bar{\e}_{(i}\,\xi_{jk)}\,^{\hat{\imath}}
      \nonumber\\[.1in]
      \delta_Q\,K_{a\,ijk}\,^{\hat{\imath}} &=&
      -\fr15\,i\,\bar{\e}^m\,(\,5\,\dslash\gamma_a+2\,\der_a\,)\,\Sigma_{mijk}\,^{\hat{\imath}}
      -\fr12\,i\,\ve_{m(i}\,\bar{\e}^m\,\gamma_a\,\xi_{jk)}\,^{\hat{\imath}}
      \nonumber\\[.1in]
      & &
      -\fr15\,i\,\ve_{ip}\,\ve_{jq}\,\ve_{kr}\,\ve^{\hat{\imath}\hat{\jmath}}\,\bar{\e}_m\,
      (\,5\,\dslash\gamma_a+2\,\der_a\,)\,\Sigma^{mpqr}\,_{\hat{\jmath}}
      \nonumber\\[.1in]
      & &
      -\fr12\,i\,\ve_{ip}\,\ve_{jq}\,\ve_{kr}\,\ve^{\hat{\imath}\hat{\jmath}}\,\ve^{m(p}\,
      \bar{\e}_m\,\gamma_a\,\xi^{qr)}\,_{\hat{\jmath}}
      \nonumber\\[.1in]
      \delta_Q\,\xi_{ij}\,^{\hat{\imath}} &=&
      \fr32\,\dslash P_{ijm}\,^{\hat{\imath}}\,\e^m
      +\fr12\,\ve^{mn}\,(\,3\,\dslash\gamma^a+8\,\der^a\,)\,K_{a\,ijm}\,^{\hat{\imath}}\,\e_n
      +\der^aA_{a\,(i}\,^{\hat{\imath}}\,\e_{j)} \,.
 \label{retmrules}\err
  These transformation rules represent the ${\cal N} =2$ supersymmetry algebra
 on all fields, without a central charge.
 The Relaxed Extended Tensor multiplet amends the Extended Tensor
 multiplet by removing the divergence-free constraint on
 $A_{a\,i}\,^{\hat{\imath}}$, and does so by amalgamating this
 multiplet with the Sextet multiplet.

 The action for the Extended Tensor multiplet given in (\ref{extenac}) is not invariant
 if we allow $\der^aA_{a\,i}\,^{\hat{\imath}}\ne 0$.  One must add
 several terms which together restore supersymmetry in the classical
 action, rendering the resulting combined action invariant under (\ref{retmrules}).
 The construction obtained in this way resolves the issue of the
 divergence-free vector, but not without cost.  On the one hand
 this introduces more ostensibly propagating fields, namely the
 bosons $Z_{ijklm}\,^{\hat{\imath}}$ and the fermions
 $\Sigma_{ijkl}\,^{\hat{\imath}}$.  But the corresponding action has deeper
 problems.  A thorough analysis, in which a
 proper basis is identified by field redefinitions, shows that
 the redefined fields $(\,\phi_i\,^{\hat{\imath}}\,|\,\psi^{\hat{\imath}}\,)$
 contribute to the Hamiltonian with opposite signs as compared to
 the contributions from
 $(\,u_{ijk}\,^{\hat{\imath}}\,,\,Z_{ijklm}\,^{\hat{\imath}}\,|\,
 \lambda_{ij}\,^{\hat{\imath}}\,,\,\Sigma_{ijkl}\,^{\hat{\imath}}\,)$.
 This does not represent a mathematical inconsistency in the classical theory.\footnote{This circumstance is analogous to the similarly
problematic
 action obtained by coupling two chiral ${\cal N} =1$ superfields $\Phi_1$ and
 $\Phi_2$ using $\int
 d^4\theta\,(\,\Phi_1^\dagger\Phi_1-\Phi_2^\dagger\Phi_2\,)$,
 resulting in a model which exhibits supersymmetry invariance at the
 classical level but is plagued with non-unitarity at the quantum
 level.} However, the putative self-action associated with (\ref{retmrules})
 thereby presents a problematic obstruction to consistent
 quantization.

 The theory is improved significantly by coupling
 a Dual Sextet multiplet as a Lagrange multiplier.  This multiplet
 has the following transformation rules,
 \brr \delta_Q\,H_i\,^{\hat{\imath}} &=&
      i\,\bar{\e}^j\,\Omega_{ij}\,^{\hat{\imath}}
      +i\,\ve_{ij}\,\ve^{\hat{\imath}\hat{\jmath}}\,\bar{\e}_k\,\Omega^{jk}\,_{\hat{\jmath}}
      \nonumber\\[.1in]
      \delta_Q\,\Omega_{ij}\,^{\hat{\imath}} &=&
      \fr43\,\dslash H_{(i}\,^{\hat{\imath}}\,\e_{j)}
      +\ve^{mn}\,T_{a\,ijm}\,^{\hat{\imath}}\,\gamma^a\,\e_n
      +R_{ijm}\,^{\hat{\imath}}\,\e^m
      \nonumber\\[.1in]
      \delta_Q\,R_{ijk}\,^{\hat{\imath}} &=&
      \fr32\,i\,\bar{\e}_{(i}\,\dslash\,\Omega_{jk)}\,^{\hat{\imath}}
      +i\,\e^{mn}\,\bar{\e}_m\,\eta_{nijk}\,^{\hat{\imath}}
      \nonumber\\[.1in]
      \delta_Q\,T_{a\,ijk}\,^{\hat{\imath}} &=&
      \fr14\,i\,\ve_{m(i}\,\bar{\e}^m(\,3\,\dslash\gamma_a-2\,\der_a\,)\,\Omega_{jk)}\,^{\hat{\imath}}
      +\fr12\,i\,\bar{\e}^m\,\gamma_a\,\eta_{mijk}\,^{\hat{\imath}}
      \nonumber\\[.1in]
      & &
      +\fr14\,i\,\ve_{ip}\,\ve_{jq}\,\ve_{kr}\,\ve^{\hat{\imath}\hat{\jmath}}\,\ve^{m(p}\,
      \bar{\e}_m\,(\,3\,\dslash\gamma_a-2\,\der_a\,)\,\Omega^{qr)}\,_{\hat{\jmath}}
       \nonumber\\[.1in]
      & &
      +\fr12\,i\,\ve_{ip}\,\ve_{jq}\,\ve_{kr}\,\ve^{\hat{\imath}\hat{\jmath}}\,\bar{\e}_m\,\gamma_a\,\eta^{mpqr}\,_{\hat{\jmath}}
      \nonumber\\[.1in]
      \delta_Q\,\eta_{ijkl}\,^{\hat{\imath}} &=&
      2\,\ve_{m(i}\,\dslash R_{jkl)}\,^{\hat{\imath}}\,\e^m
      +\fr25\,(\,5\,\dslash\gamma^a+8\,\der^a\,)\,T_{a\,(ijk}\,^{\hat{\imath}}\,\e_{l)}
      +\ve^{mn}\,Y_{ijklm}\,^{\hat{\imath}}\,\e_n
      \nonumber\\[.1in]
      \delta_Q\,Y_{ijklm}\,^{\hat{\imath}} &=&
      -2\,i\,\ve_{n(i}\,\bar{\e}^n\,\dslash\eta_{jklm)}\,^{\hat{\imath}}
      -2\,i\,\ve_{ip}\,\ve_{jq}\,\ve_{kr}\,\ve_{ls}\,\ve_{mt}\,\ve^{\hat{\imath}\hat{\jmath}}\,\ve^{n(p}\,
      \bar{\e}_n\,\dslash\eta^{qrst)}\,_{\hat{\jmath}} \,.
 \label{dualsex}\err
 Note that the highest component $Y_{ijklm}\,^{\hat{\imath}}$ has the appropriate
 $SU(2)_R\times SU(2)_H$ structure to couple invariantly to the field $Z_{ijklm}\,^{\hat{\imath}}$,
 so as to render $Z_{ijklm}\,^{\hat{\imath}}$ auxiliary.  This can
 be done supersymmetrically; indeed it is this feature which gives
 this multiplet its name. As an extra bonus, the modified action has a healthy kinetic
 sector in the sense that the Hamiltonian is properly positive
 definite, so as not to preclude quantization.

\begin{figure}
 \begin{center}
 \includegraphics[width=5in]{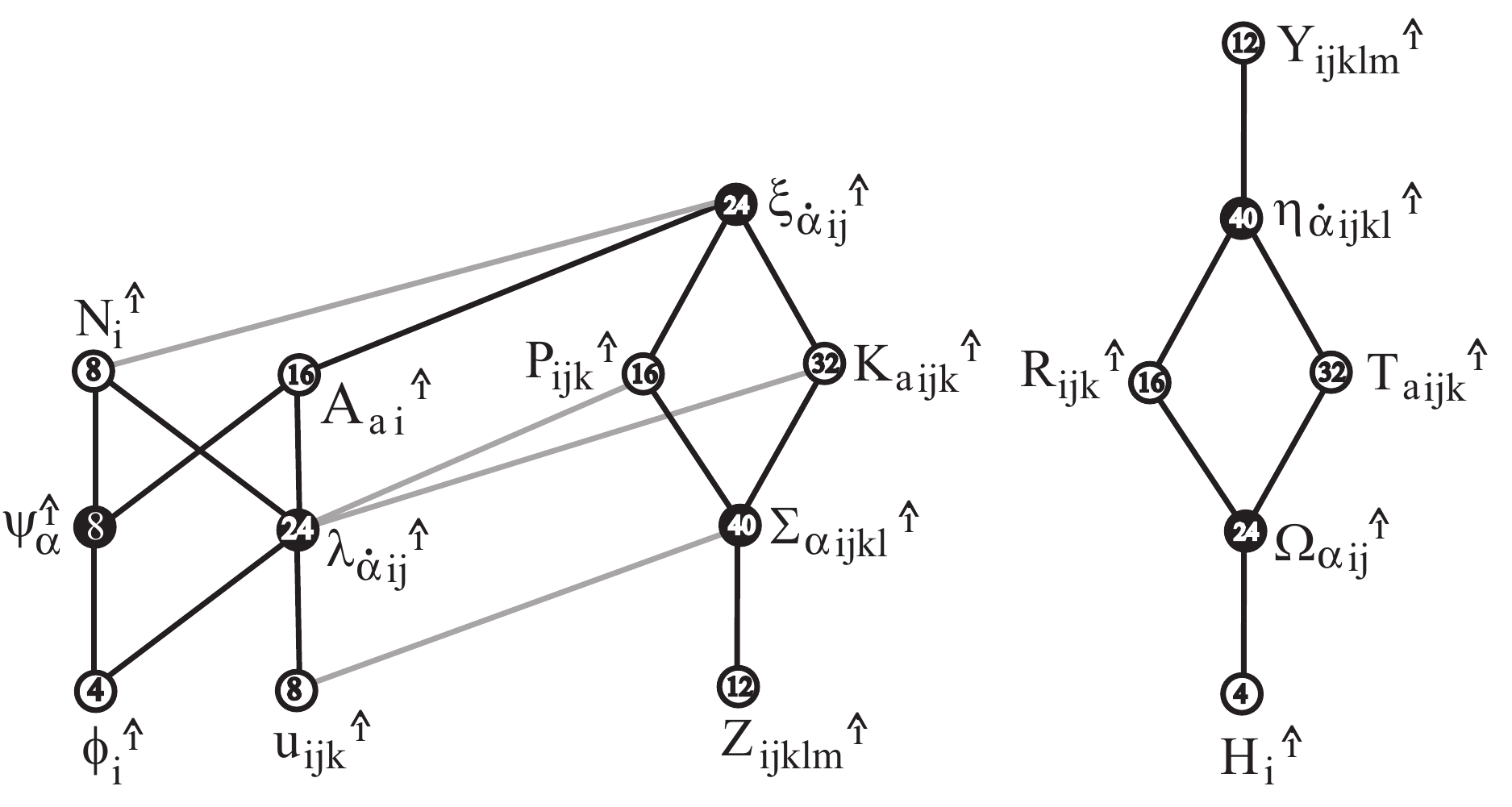}
 \caption{Adinkraic depiction of the transformation rules given
 in (\ref{retmrules}) and (\ref{dualsex}).}
 \label{retmadinkras}
 \end{center}
 \end{figure}

 The structure of this theory is rendered helpfully perspicuous
 using Adinkra diagrams \cite{FG1,DFGHIL01,DFGHIL00,Boats}, as shown in Figure \ref{retmadinkras}.
 These represent the transformation rules graphically;
 white nodes correspond to off-shell bosons and black nodes
 correspond to off-shell fermions. The state counting is
 indicated by the numerals in the nodes, and inter-node edge
 connections represent terms in the supersymmetry transformation
 rules.  The vertical placement of the nodes correspond faithfully to
 the relative engineering dimensions of the corresponding fields; ``lower" nodes
 correspond to fields with lower engineering dimension while ``higher" nodes
 have greater engineering dimension.  Black edges correspond to terms which act both ``upward", by
 transforming fields with a lower engineering dimension into
 fields with higher engineering dimension, and also ``downward" by transforming fields with a
 higher engineering dimension into derivatives of fields with a lower engineering dimension.  Grey edges correspond to
 terms which only act ``upward".  It is a noteworthy fact about the
 Relaxed Extended Tensor multiplet that its Adinkra contains grey
 edges.

 The supersymmetric action corresponding to the Relaxed Extended Tensor multiplet
 (RETM) coupled to a Dual Sextet multiplet is given by
 \brr S &=& \int d^4x\,\bpl
      -\fr34\,\phi_i\,^{\hat{\imath}}\,\Box\,\phi^i\,_{\hat{\imath}}
      -\fr92\,u_{ijk}\,^{\hat{\imath}}\,\Box\,u^{ijk}\,_{\hat{\imath}}
      +\fr{32}{3}\,Z^{ijklm}\,_{\hat{\imath}}\,\Box\,Z_{ijklm}\,^{\hat{\imath}}
      \nonumber\\[.1in]
      & & \hspace{.7in}
      +N_i\,^{\hat{\imath}}\,N^i\,_{\hat{\imath}}
      -A_{a\,i}\,^{\hat{\imath}}\,A^{a\,i}\,_{\hat{\imath}}
      -\fr{16}{3}\,P^{ijk}\,_{\hat{\imath}}\,P_{ijk}\,^{\hat{\imath}}
      +\fr{16}{3}\,K_a^{ijk}\,_{\hat{\imath}}\,K_{a\,ijk}\,^{\hat{\imath}}
      \nonumber\\[.1in]
      & & \hspace{.7in}
      -i\,\bar{\psi}^{\hat{\imath}}\dslash \psi_{\hat{\imath}}
      -\fr92\,i\,\bar{\lambda}_{ij}\,^{\hat{\imath}}\,\dslash\lambda^{ij}\,_{\hat{\imath}}
      +\fr{32}{3}\,i\,\bar{\Sigma}^{ijkl}\,_{\hat{\imath}}\,\dslash\Sigma_{ijkl}\,^{\hat{\imath}}
      \nonumber\\[.1in]
      & & \hspace{.7in}
      +4\,i\,\ve^{im}\,\ve^{jn}\,\ve_{\hat{\imath}\hat{\jmath}}\,
      \bar{\xi}_{ij}\,^{\hat{\imath}}\,\lambda_{mn}\,^{\hat{\jmath}}
      +4\,i\,\ve_{im}\,\ve_{jn}\,\ve^{\hat{\imath}\hat{\jmath}}\,
      \bar{\xi}^{ij}\,_{\hat{\imath}}\,\lambda^{mn}\,_{\hat{\jmath}}
      \nonumber\\[.1in]
      & & \hspace{.7in}
      +A_{a\,i}\,^{\hat{\imath}}\,\der^a\phi^i\,_{\hat{\imath}}
      +16\,K_a^{ijk}\,_{\hat{\imath}}\,\der^au_{ijk}\,^{\hat{\imath}}
      -Z_{ijklm}\,^{\hat{\imath}}\,Y^{ijklm}\,_{\hat{\imath}}
      \nonumber\\[.1in]
      & & \hspace{.7in}
      +H^i\,_{\hat{\imath}}\,\der^aA_{a\,i}\,^{\hat{\imath}}
      -2\,K_a^{ijk}\,_{\hat{\imath}}\,T^a_{ijk}\,^{\hat{\imath}}
      +P_{ijk}\,^{\hat{\imath}}\,R^{ijk}\,_{\hat{\imath}}
      +P^{ijk}\,_{\hat{\imath}}\,R_{ijk}\,^{\hat{\imath}}
      \nonumber\\[.1in]
      & & \hspace{.6in}
      -i\,\bar{\eta}^{ijkl}\,_{\hat{\imath}}\,\Sigma_{ijkl}\,^{\hat{\imath}}
      -i\,\bar{\eta}_{ijkl}\,^{\hat{\imath}}\,\Sigma^{ijkl}\,_{\hat{\imath}}
      -i\,\bar{\xi}^{ij}\,_{\hat{\imath}}\,\Omega_{ij}\,^{\hat{\imath}}
      -i\,\bar{\xi}_{ij}\,^{\hat{\imath}}\,\Omega^{ij}\,_{\hat{\imath}}\,\bpr
      \,.
 \label{rsaction}\err
 This action is invariant under the transformation rules
 given in (\ref{retmrules}) and (\ref{dualsex}).
 The first fourteen terms are invariant by themselves, and correspond
 to the unhealthy RETM action referred to above.  The remaining nine terms are also
 invariant by themselves, and correspond to the invariant coupling of the Dual
 Sextet multiplet to the Sextet portion of the RETM.  The addition
 of these extra terms has two significant effects, one to render the
 boson  $Z_{ijklm}\,^{\hat{\imath}}$ and the fermion
 $\Sigma_{ijkl}\,^{\hat{\imath}}$ auxiliary, and the other to
 repair the positivity of the Hamiltonian.

 To find a canonical basis, we redefine the component fields in
 (\ref{rsaction}) according to
 \brr A_{a\,i}\,^{\hat{\imath}} &\to&
      A_{a\,i}\,^{\hat{\imath}}-\fr12\,H_i\,^{\hat{\imath}}
      \nonumber\\[.1in]
      H_i\,^{\hat{\imath}} &\to&
      H_i\,^{\hat{\imath}}
      +\phi_i\,^{\hat{\imath}}
      \nonumber\\[.1in]
      P_{ijk}\,^{\hat{\imath}} &\to&
      P_{ijk}\,^{\hat{\imath}}
      +\fr{3}{16}\,R_{ijk}\,^{\hat{\imath}}
      \nonumber\\[.1in]
      K_{a\,ijk}\,^{\hat{\imath}} &\to&
      K_{a\,ijk}\,^{\hat{\imath}}
      +\fr{3}{16}\,T_{a\,ijk}\,^{\hat{\imath}}
      \nonumber\\[.1in]
      T_{a\,ijk}\,^{\hat{\imath}} &\to&
      T_{a\,ijk}\,^{\hat{\imath}}
      +8\,\der_au_{ijk}\,^{\hat{\imath}}
      \nonumber\\[.1in]
      \Omega_{\alpha\,ij}\,^{\hat{\imath}} &\to&
      \Omega_{\alpha\,ij}\,^{\hat{\imath}}
      +4\,\ve_{im}\,\ve_{jn}\,\ve^{\hat{\imath}\hat{\jmath}}\,\lambda^{mn}\,_{\hat{\jmath}}
      \nonumber\\[.1in]
      \eta_{ijkl}\,^{\hat{\imath}} &\to&
      \eta_{ijkl}\,^{\hat{\imath}}
      +\fr{16}{3}\,\dslash\Sigma_{ijkl}\,^{\hat{\imath}}
      \nonumber\\[.1in]
      Y_{ijklm}\,^{\hat{\imath}} &\to&
      Y_{ijklm}\,^{\hat{\imath}}
      +\fr{32}{3}\,\Box\,Z_{ijklm}\,^{\hat{\imath}} \,.
 \err
 When expressed in terms of the re-defined fields, the action (\ref{rsaction}) becomes
  \brr S &=& \int d^4x\,\bpl
      -\fr34\,\phi_i\,^{\hat{\imath}}\,\Box\,\phi^i\,_{\hat{\imath}}
      -\fr92\,u_{ijk}\,^{\hat{\imath}}\,\Box\,u^{ijk}\,_{\hat{\imath}}
      -\fr14\,H^i\,_{\hat{\imath}}\,\Box\,H_i\,^{\hat{\imath}}
      \nonumber\\[.1in]
      & & \hspace{.7in}
      -i\,\bar{\psi}^{\hat{\imath}}\dslash \psi_{\hat{\imath}}
      -\fr92\,i\,\bar{\lambda}_{ij}\,^{\hat{\imath}}\,\dslash\lambda^{ij}\,_{\hat{\imath}}
      \nonumber\\[.1in]
      & & \hspace{.7in}
      -i\,\bar{\eta}^{ijkl}\,_{\hat{\imath}}\,\Sigma_{ijkl}\,^{\hat{\imath}}
      -i\,\bar{\eta}_{ijkl}\,^{\hat{\imath}}\,\Sigma^{ijkl}\,_{\hat{\imath}}
      -i\,\bar{\xi}^{ij}\,_{\hat{\imath}}\,\Omega_{ij}\,^{\hat{\imath}}
      -i\,\bar{\xi}_{ij}\,^{\hat{\imath}}\,\Omega^{ij}\,_{\hat{\imath}}
      \nonumber\\[.1in]
      & & \hspace{.7in}
      +\fr{16}{3}\,K_a^{ijk}\,_{\hat{\imath}}\,K_{a\,ijk}\,^{\hat{\imath}}
      -\fr{3}{16}\,T_a^{ijk}\,_{\hat{\imath}}\,T^a_{ijk}\,^{\hat{\imath}}
      -A_{a\,i}\,^{\hat{\imath}}\,A^{a\,i}\,_{\hat{\imath}}
      -Z_{ijklm}\,^{\hat{\imath}}\,Y^{ijklm}\,_{\hat{\imath}}
      \nonumber\\[.1in]
      & & \hspace{.7in}
      -\fr{16}{3}\,P^{ijk}\,_{\hat{\imath}}\,P_{ijk}\,^{\hat{\imath}}
      -\fr{3}{16}\,R^{ijk}\,_{\hat{\imath}}\,R_{ijk}\,^{\hat{\imath}}
      +N_i\,^{\hat{\imath}}\,N^i\,_{\hat{\imath}}\,\bpr \,.
 \label{shifted}\err
 This action describes 96+96 degrees of freedom off-shell, and 16+16
 degrees of freedom on-shell.  The propagating fields are given by
 $(\,\phi_i\,^{\hat{\imath}}\,,\,u_{ijk}\,^{\hat{\imath}}\,,\,H_i\,^{\hat{\imath}}\,|\,
 \psi^{\hat{\imath}}\,,\,\lambda_{ij}\,^{\hat{\imath}}\,)$.
 Notice that all kinetic terms in (\ref{shifted}) have consistent signs.

 The model described by (\ref{shifted}) has ${\cal N} =2$ supersymmetry with no central charge,
 finite off-shell degrees of freedom, and a Hypermultiplet sector,
 the latter comprising a pairing of boson fields and fermion fields transforming under
 $SU(2)_R\times SU(2)_H$ with the representation content implicit in
 $\phi_i\,^{\hat{\imath}}$ and $\psi^{\hat{\imath}}$.  We believe
 this represents the first appearance of such an action, and this
 is the principal result motivating this paper.

 We have done further analysis on the structure described above, and
 have identified a compelling extension to this model.  There exists
 a multiplet, namely the Dual to a constrained Quadruplet multiplet, which has the proper field content to render the
 fields $u_{ijk}\,^{\hat{\imath}}$ and $\lambda_{ij}\,^{\hat{\imath}}$
 non-propagating while supplying solely an additional Weyl spinor
 $SU(2)_H$ doublet, which we call $\beta^{\hat{\imath}}$, to the set of propagating fields.  Thus, in this
 case the ostensibly propagating fields would have precisely the content of two
 Hypermultiplets, $(\,\phi_i\,^{\hat{\imath}}\,|\,\psi^{\hat{\imath}}\,)$ and
 $(\,H_i\,^{\hat{\imath}}\,|\,\beta^{\hat{\imath}}\,)$.  However,
 the action obtained in this way has the curious feature that the
 two Hypermultiplet sectors contribute with opposite signs to the
 Hamiltonian.

\begin{figure}
 \begin{center}
 \includegraphics[width=6in]{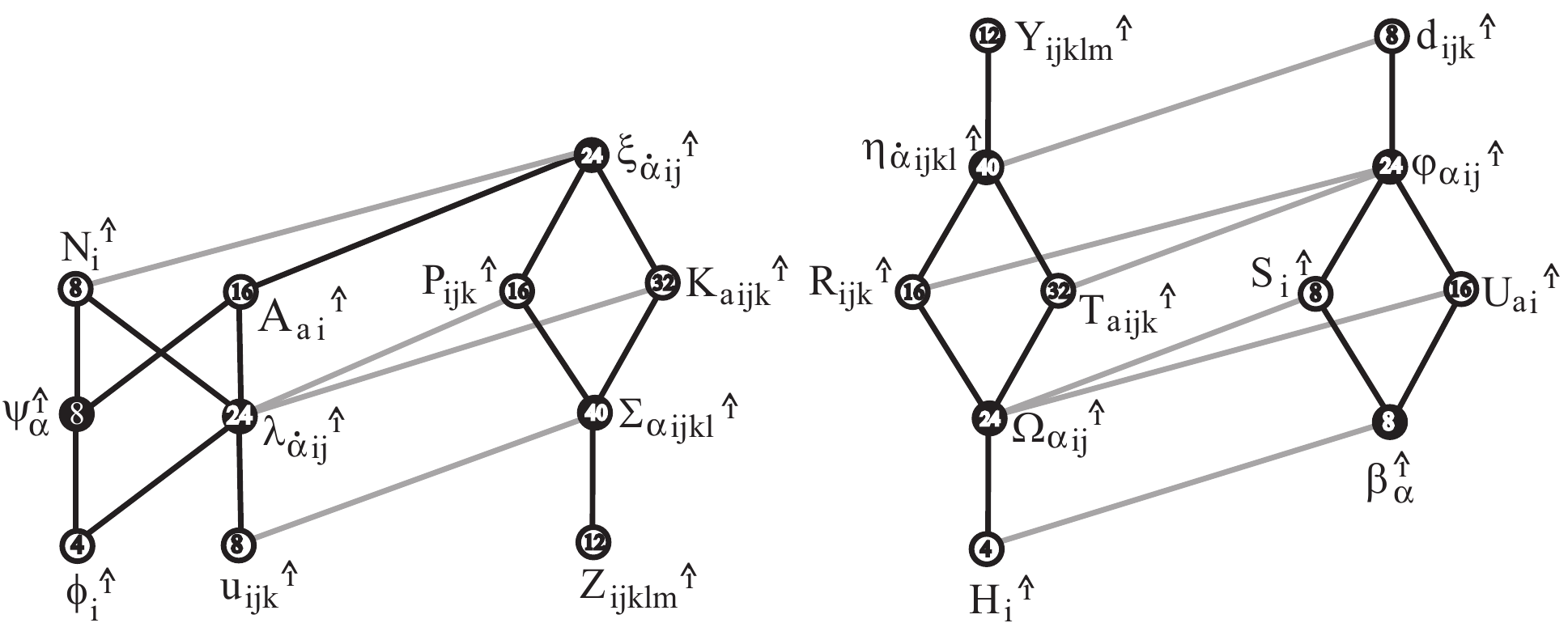}
 \caption{The Releaxed Extended Tensor multiplet admits, at the classical level,
 a coupling to a particular amalgamation of the Dual Sextet multiplet with the
 Dual Quadruplet multiplet, the latter of which is represented by the right-most
 agglomeration in the above Adinkra.}
 \label{retmdqdsa}
 \end{center}
 \end{figure}

 The particular way in which the Dual Quadruplet couples to the
 Relaxed Extended Tensor multiplet is interesting.  A supersymmetric invariant action
 involving these fields exists only if the
 transformation rules for the Dual Sextet multiplet are extended to
 admit ``one-way" terms in which Dual Sextet fields transform into
 Dual Quadruplet fields, but not the other way around.  These terms
 are codified by the grey edges in the right-hand connected
 component of the Adinkra representation of this extended model
 shown in Figure \ref{retmdqdsa}.  These represent a
 ``quasi-relaxation" which closely mimics the one-way terms present in the
 Relaxed Extended Tensor multiplet.  The off-shell states comprising
 the Dual Quadruplet multiplet are also clarified in Figure
 \ref{retmdqdsa}.

 It is conceivable that by coupling the above model to ${\cal N} =2$ Vector multiplets
 that the canonical structure of the resulting action would repair the non-unitarity
 present in the system codified in Figure \ref{retmdqdsa}, thereby producing a perfectly
 consistent model involving two Hypermultiplets coupled to Vector
 multiplets, having
 finite off-shell degrees of freedom and no off-shell central charge.  It is also possible that
 such a structure would split into separable Hypermultiplet-Vector multiplet couplings, one or
 both of which could be consistently removed, thereby revealing an off-shell representation
 involving only one Hypermultiplet and only one Vector multiplet.
 In either case, the elucidation of such a structure would likely
 shed surprising light on a possible off-shell extension to the
 important case of ${\cal N} =4$ Super Yang Mills theory, an application
 which has been a prime motivator throughout our work.  We are
 currently engaged in developing these ideas further.  But, we find
 the mere existence of the off-shell extension of the dual-Hypermultiplet pairing
 exhibited in Figure \ref{retmdqdsa} surprising and noteworthy in
 its own right, regardless of what mechanisms might or might not
 exist to cure its non-unitarity.

  As another application of these ideas, the models described in this paper should generalize to
 provide non-linear sigma models with various uses.
 For example, these should allow a new class of c-maps
 \cite{CMAPS}; as known since 1984, by reducing
 four-dimensional field theories involving vector or tensor fields to two dimensions
 one obtains new non-linear sigma-models.  This follows because the two-dimensional analog of the condition
 $\der^aA_{a\,i}\,^{\hat{\imath}} = 0$  possesses
 the solution $\der^aA_{a\,i}\,^{\hat{\imath}} $ $=$ $\epsilon_{a \,
 b}\,\der^b \varphi_{\,i}\,^{\hat{\imath}}$.  In particular, the action (\ref{extenac}) reduces
 in two-dimensions to a model having off-shell 2d, $\cal N$ = 4 supersymmetry.
 It follows that this
 must possess non-linear $\sigma$-model extensions that lead to a
 proper c-map.  Such non-linear $\sigma$-models are characterized by
 possessing a K\" ahler-like prepotential that is a doublet-doublet
 under $SU(2)_R \times SU(2)_H$, or a doublet-$2\,n$-plet under $SU(2)_H$ $\to$ $Sp(2n)$ for $n$ $>$
 1. It is also interesting to note that under a reduction to two dimensions,
 the dynamical sectors of (\ref{extenac}) and (\ref{shifted}) become
 identical.

 In conclusion, we have exhibited a concrete example of an off-shell
 extension of a Hypermultiplet $(\,\phi_i\,^{\hat{\imath}}\,|\,\psi^{\hat{\imath}}\,)$
 coupled to ``Quadruplet" matter
 $(\,u_{ijk}\,^{\hat{\imath}}\,|\,\lambda_{ij}\,^{\hat{\imath}}\,)$,
 which has finite off-shell degrees of freedom and no off-shell
 central charge.  We have elucidated some novel forms of ${\cal N} =2$
 matter, namely the Extended Tensor multiplet, the Sextet multiplet, the Relaxed Extended Tensor multiplet,
 the Dual Sextet multiplet, the Dual Quadruplet multiplet, and some interesting supersymmetric couplings
 involving these.  We have also described the rudiments of a
 paradigm based on these constructions which seems to suggest a way
 to understand features of off-shell extensions to ${\cal N} =4$ Super
 Yang-Mills theory. \newline $~~~$ \newline

\noindent
{\Large\bf Acknowledgments}

This research has been supported in part by NSF Grant PHY-06-52363,
the J.~S. Toll Professorship endowment and the UMCP Center for
String \& Particle Theory. T.~H. is indebted to the generous support
of the Department of Energy
  through the grant DE-FG02-94ER-40854.
  C.~F.~D. was supported in part by the University of Washington Royalty Research
  Fund.  M.~F. expresses much gratitude to the Slovak Institute for Basic Research, Podva$\check{\rm z}$ie,
  Slovakia, where much of this work was completed.

  \Refs{References}{[00]}
 \Bib{Fayet}
 P.~Fayet,
 {\it Fermi-Bose Hypersymmetry},
  Nucl.\ Phys.\ B {\bf 113}, 135 (1976);
  \Bib{Sohnius}
  M.~F.~Sohnius,
  {\it Supersymmetry and Central Charges},
  Nucl.\ Phys.\ B {\bf 138}, 109 (1978);
  \Bib{Harmonic}
  A.~S.~Galperin, E.~A.~Ivanov, V.~I.~Ogievetsky, and
  E.~S.~Sokatchev,
  {\it Harmonic Superspace},
  Cambridge University Press (2001);
  \Bib{Tensors}
  W.~Siegel,
  {\it Gauge Spinor Superfield as a Scalar Multiplet},
  Phys.\ Lett.\ B85 (1979) 333;
 \Bib{HST}
  P.~S.~Howe, K.~S.~Stelle and P.~K.~Townsend,
  {\it The Relaxed Hypermultiplet: An Unconstrained ${\cal N}=2$ Superfield
  Theory},
  Nucl.\ Phys.\ B {\bf 214}, 519 (1983);
 \Bib{FG1} M.~Faux and S.~J.~Gates, Jr.:
 {\em Adinkras: A Graphical Technology for Supersymmetric Representation Theory},
  Phys.~Rev.~{\bf D71}~(2005),~065002;
  \Bib{DFGHIL01} C.~Doran, M.~Faux, S.~J.~Gates, Jr., T.~H{\"u}bsch, K.~Iga, G.~Landweber:
 {\em On Graph Theoretic Identifications of
 Adinkras, Supersymmetry Representations and Superfields},
  math-ph/0512016;
 \Bib{DFGHIL00} C.~Doran, M.~Faux, S.~J.~Gates, Jr., T.~H{\"u}bsch, K.~Iga, G.~Landweber:
 {\em  Off-Shell Supersymmetry and Filtered Clifford Supermodules},
  math-ph/0603012;
  \Bib{Prepotentials}
  C.~Doran, M.~Faux, S.~J.~Gates, Jr., T.~H{\"u}bsch, K.~Iga, G.~Landweber:
  {\em Adinkras and the Dynamics of Superspace Prepotentials},
  Adv. S. Th. Phys. (in press), hep-th/0605269;
  \Bib{Boats}
  C.~Doran, M.~Faux, S.~J.~Gates, Jr., T.~H{\"u}bsch, Jr., K.~Iga and
  G.~D.~Landweber:
  {\em A Counter-Example to a Putative Classification of
  1-Dimensional, N-Extended Supermultiplets},
   Adv. S. Th. Phys. (in press), hep-th/0611060;
 \Bib{CMAPS}
  S.~J.~Gates, Jr., Nucl. Phys. B238 (1984) 349,
 S. Cecotti, S. Ferrara, and L. Girardello, In. J. Mod. Phys. A4
 (1989) 2475.
 \endRefs

 \end{document}